\documentclass{moriond}

\def\be{\begin{equation}}
\def\ee{\end{equation}}
\def\bea{\begin{eqnarray}}
\def\eea{\end{eqnarray}}

\newcommand{\Photo}{}

\begin{document}
\vspace*{4cm}
\title{Forward-Backward Drell-Yan Asymmetry and PDF 
Determination~\footnote{Presented by F Hautmann at the 54th Rencontres de Moriond, La Thuile, March 2019.}}

\author{H Abdolmaleki, E Accomando, V Bertone, J Fiaschi, F Giuli, A Glazov,\\ 
 F Hautmann, A Luszczak, S Moretti, I Novikov, F Olness, O Zenaiev}

\address{~}

\maketitle\abstracts{We 
investigate the impact of  high-statistics Drell-Yan (DY) measurements at the    LHC  on 
the study of non-perturbative QCD effects from 
 parton distribution functions (PDF).  We   present the results of a PDF profiling analysis based on the neutral-current  
DY forward-backward asymmetry,  
  using the open source fit platform {\tt{xFitter}}.
}

\vspace*{-8cm} 
\begin{flushright} 
CERN-TH-2019-116
\end{flushright} 
\vspace*{8cm}

The high statistics  at the Large Hadron Collider (LHC) Run II and the forthcoming Run III and  high-luminosity  HL-LHC 
open the way to precision measurements at the TeV scale, which will be used both for studies of the Standard Model (SM) and for searches for beyond-Standard-Model (BSM) physics.
In order to keep up with the increasing statistical precision of experimental measurements, an impressive effort is being made on the theoretical side to provide higher-order 
 perturbative QCD calculations  --- see e.g.~\cite{Wackeroth:2019xib}. 
With improving perturbative accuracy, nonperturbative QCD contributions such as parton distribution functions (PDF) more and more become a crucial limiting factor in the theoretical systematics affecting both precision SM studies and BSM searches.
An important  part of the physics program to be carried out with current and upcoming collider data is thus to identify which measurements can be most helpful in placing constraints on the nonperturbative PDF and their uncertainties. 

In the Drell-Yan (DY)  production channel, measurements of differential distributions in  mass and rapidity and of the charged-current (CC) asymmetry have long been used to constrain PDFs (see e.g.~\cite{Aaboud:2016btc,Dulat:2015mca,Ball:2017nwa,Alekhin:2017kpj,Harland-Lang:2014zoa,Abramowicz:2015mha} for recent results), while    
measurements of the neutral current (NC) forward-backward asymmetry (henceforth denoted as $A_{\rm{FB}}$) have traditionally been 
used  for determinations of the weak mixing angle $\theta_W$  (see 
e.g.~\cite{Sirunyan:2018swq,ATLAS2018037,Bodek:2016olg,Aad:2015uau,Chatrchyan:2011ya,Aaij:2015lka}). 
In~\cite{Accomando:2018nig,Accomando:2017scx} it was observed  that   $A_{\rm{FB}}$  measurements in NC processes  at the LHC 
 can usefully be employed for PDF determinations.  Ref.~\cite{Accomando:2019}    investigates the impact of $A_{\rm{FB}}$  data on PDF extractions 
 by using the open source fit platform   {\tt{xFitter}}~\cite{Alekhin:2014irh}, considering  different scenarios  for luminosities (from 
 Runs II, III to the HL-LHC stage~\cite{Azzi:2019yne}) and performing PDF profiling to analyze quantitatively the effect of $A_{\rm{FB}}$  on PDF uncertainties.  In  
 this article we report on this study.

The five-fold differential DY cross section in the vector boson mass, rapidity,  transverse momentum and lepton decay angles
may be written in terms of angular coefficients $A_k$ as 
\begin{eqnarray} 
{ {d \sigma } \over {d M_{\ell \ell} d Y_{\ell \ell} d P^\perp_{\ell \ell} d \cos \theta  d \phi }} 
& = &  { { d \sigma^{({\rm{U}})} } \over {d M_{\ell \ell} d Y_{\ell \ell} d P^{\perp}_{\ell \ell} }} { 3 \over { 16 \pi} } 
\left[ 1 + \cos^2 \theta + {1 \over 2} A_0 ( 1 - 3 \cos^2 \theta  ) 
\right. 
\\ 
& + & A_1 \sin 2 \theta \cos \phi  +   {1 \over 2} A_2 \sin^2 \theta \cos 2 \phi + A_3 \sin \theta \cos  \phi 
\nonumber\\ 
& + &   \left.   A_4 \cos \theta + 
A_5 \sin^2 \theta \sin 2 \phi  +  A_6 \sin 2 \theta \sin  \phi  + A_7 \sin \theta \sin \phi 
\right] \;  . 
\nonumber 
\end{eqnarray} 
The azimuthally integrated cross section is  given by 
\begin{equation}  
{ {d \sigma } \over {d M_{\ell \ell} d Y_{\ell \ell} d P^\perp_{\ell \ell} d \cos \theta}} 
 =   { { d \sigma^{({\rm{U}})} } \over {d M_{\ell \ell} d Y_{\ell \ell} d P^{\perp}_{\ell \ell} }} { 3 \over {8} } 
\left[ 1 + \cos^2 \theta + {1 \over 2} A_0 ( 1 - 3 \cos^2 \theta  ) +  
A_4 \cos \theta 
 \right]  \; , 
\end{equation}  
where the $A_4$ term is responsible for the forward-backward asymmetry. This may be defined as  
\begin{equation}  
\label{equ3} 
A_{\rm{FB}} =   { { \sigma_F - \sigma_B } \over { \sigma_F + \sigma_B } }  
\;\;\;\; \;\;\;\; 
%
{\rm{where}} \;\;\;\;  \sigma_F  = \int_0^1 { {d \sigma } \over { d \cos \theta}}  \   d \cos \theta \;\; , \;\;\;\;  
\sigma_B  = \int_{-1}^0 { {d \sigma } \over { d \cos \theta}}  \   d \cos \theta  \; . 
\end{equation}  
At leading order (LO)  in $\alpha_s$,    $A_0 =0$, $A_4 \neq 0$. 
The LO triple differential cross section may be written  as 
\begin{equation}
 \frac{d \sigma}{dM_{\ell\ell}dY_{\ell\ell}d\cos\theta} = \frac{\pi\alpha^2}{3M_{\ell\ell}s} \sum_q H_q \left[f_q  f_{\bar{q}}  +  \{ q \leftrightarrow {\bar q} \}   \right]   \; , 
\end{equation}
where  $f$ is the  PDF and $H_q$ is given in terms of the vector and axial  couplings $v$ and $a$ and electric charges $e$ by
\begin{eqnarray} 
\label{equ5}
 H_q &=&  e^2_\ell e^2_q (1 + \cos^2\theta)
  \\ 
 &+& e_\ell e_q \ 
 \frac{2M^2_{\ell\ell}(M^2_{\ell\ell} - M^2_Z)}{\sin^2\theta_W \cos^2\theta_W\left[(M^2_{\ell\ell} - M^2_Z)^2 + \Gamma^2_Z M^2_Z\right]}  \left[v_\ell v_q (1 + \cos^2\theta) + 2 a_\ell a_q \cos\theta\right] 
 \nonumber\\
 &+& \frac{M^4_{\ell\ell}}{\sin^4\theta_W \cos^4\theta_W\left[(M^2_{\ell\ell} - M^2_Z)^2 + \Gamma^2_Z M^2_Z\right]} 
 \nonumber\\
&\times&  [(a^2_\ell + v^2_\ell) (a^2_q + v^2_q) (1+\cos^2\theta)
 + 8 a_\ell v_\ell a_q v_q \cos\theta] \; . 
  \nonumber
\end{eqnarray}
The  $ A_{\rm{FB}} $ is dominated by the $ Z / \gamma $ interference $\cos \theta$ term in the second line of 
Eq.~(\ref{equ5}), proportional to $ e_\ell e_q   a_\ell a_q $, with $ a_q = T_q^3 / 2 $, where $T_q^3$ is the third component of weak isospin.  
It is thus primarily  sensitive  to  the charge-weighted PDF linear combination $ (2 / 3) u + (1 / 3) d $.  

To carry out the PDF profiling analysis, the  $ A_{\rm{FB}} $ is implemented in  {\tt{xFitter}}, and 
  NLO QCD corrections to DY  are included via NLO grids obtained with  {\tt{MadGraph5{\_}aMC@NLO}}~\cite{Alwall:2014hca}, 
  interfaced to {\tt{APPLgrid}}~\cite{Carli:2010rw} through {\tt{aMCfast}}~\cite{Bertone:2014zva}. 
     At LO  the angle $\theta$ may be reconstructed using the direction of the boost of the di-lepton 
  system~\cite{Dittmar:1996my,Rizzo:2009pu,Accomando:2016tah,Accomando:2016ehi,Accomando:2015cfa}, while in general we use the definition of angle 
  $\theta$ in the 
  CS frame~\cite{Collins:1977iv}.  The cross sections are computed in the detector fiducial region using the  acceptance cuts of~\cite{Aaboud:2017ffb}. 
  Suitable datafiles with pseudodata are generated for the profiling analysis as described in~\cite{Accomando:2019}.

\begin{figure}[h]
\begin{center}
\includegraphics[width=0.33\textwidth]{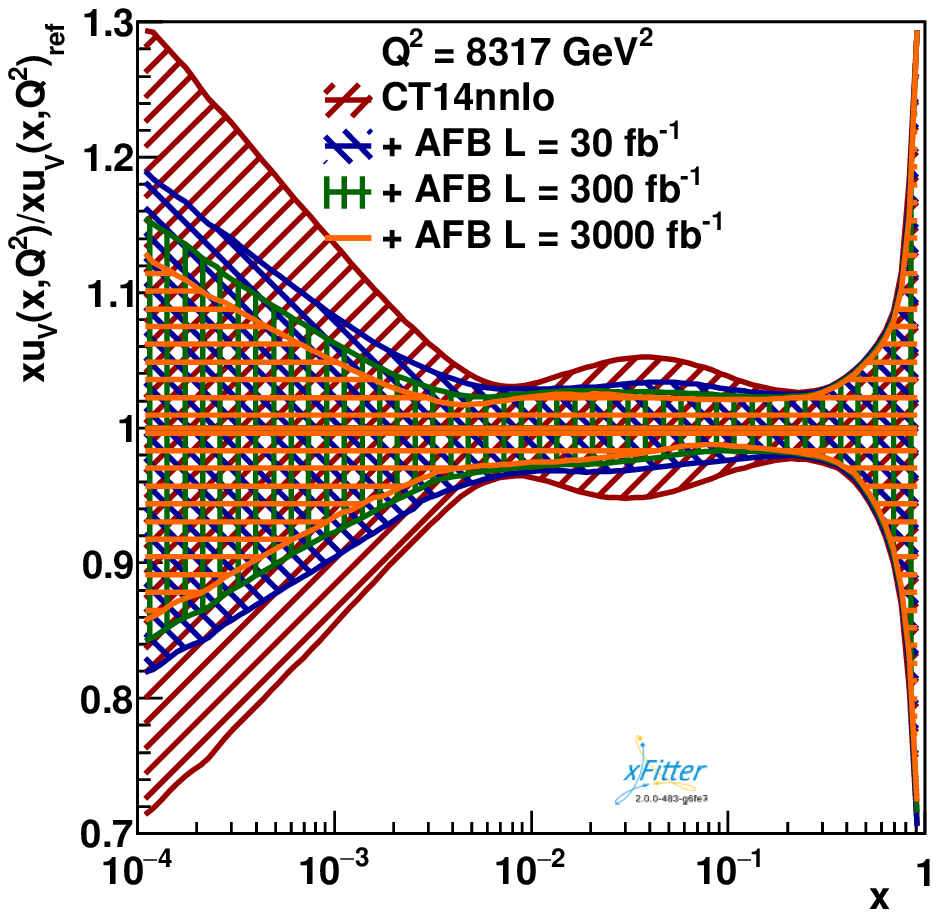}%
\includegraphics[width=0.33\textwidth]{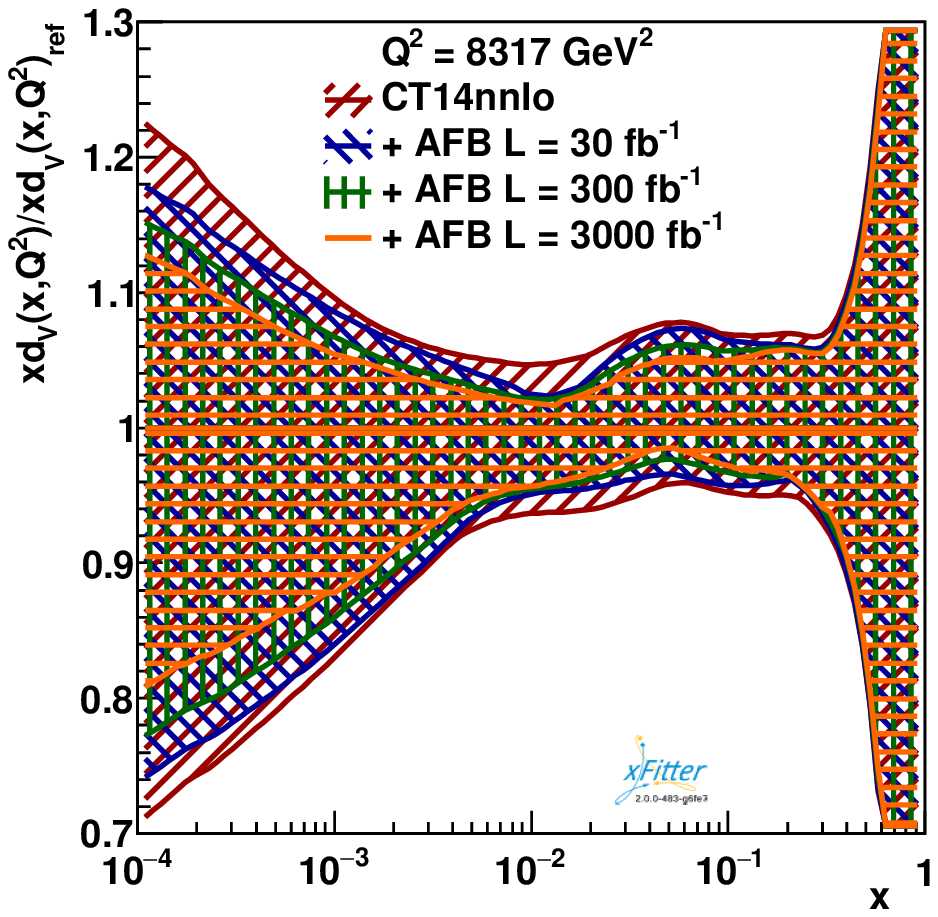}%
\includegraphics[width=0.33\textwidth]{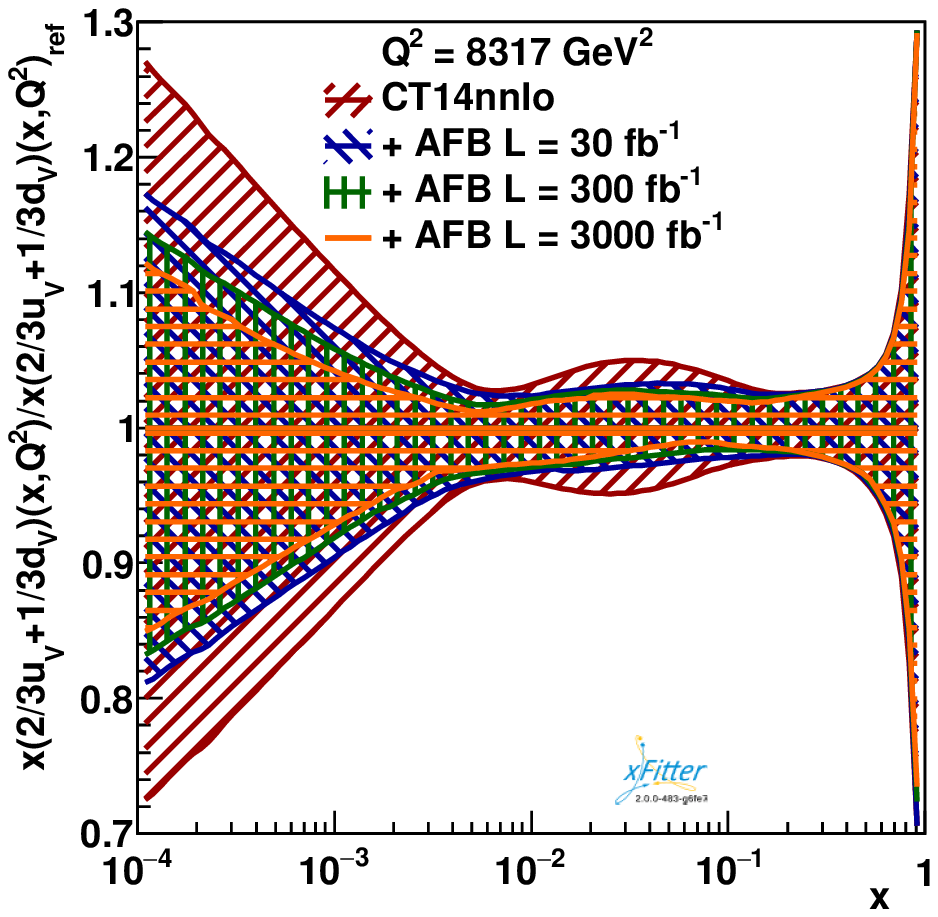}
\caption{Original (red) and profiled curves distributions for the normalised distribution of the ratios of (left to right) $u$-valence, $d$-valence and $((2/3)u+(1/3)d)$-valence 
of the CT14nnlo PDF set,  using $A_{\rm{FB}}$ pseudodata corresponding to  integrated luminosities of 30 fb$^{-1}$ (blue), 300 fb$^{-1}$ (green) and 3000 fb$^{-1}$ (orange).}
\label{fig:prof_CT14nnlo_lum}
\end{center}
\end{figure}

\begin{figure}[h]
\begin{center}
\includegraphics[width=0.25\textwidth]{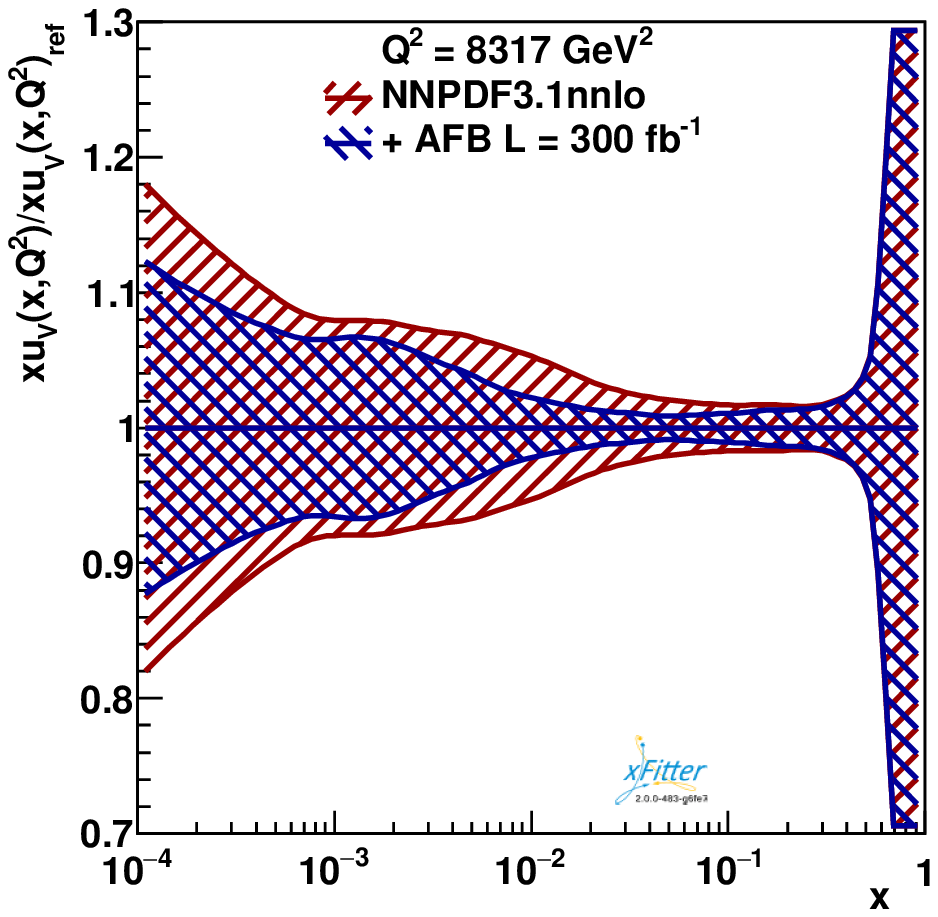}%
\includegraphics[width=0.25\textwidth]{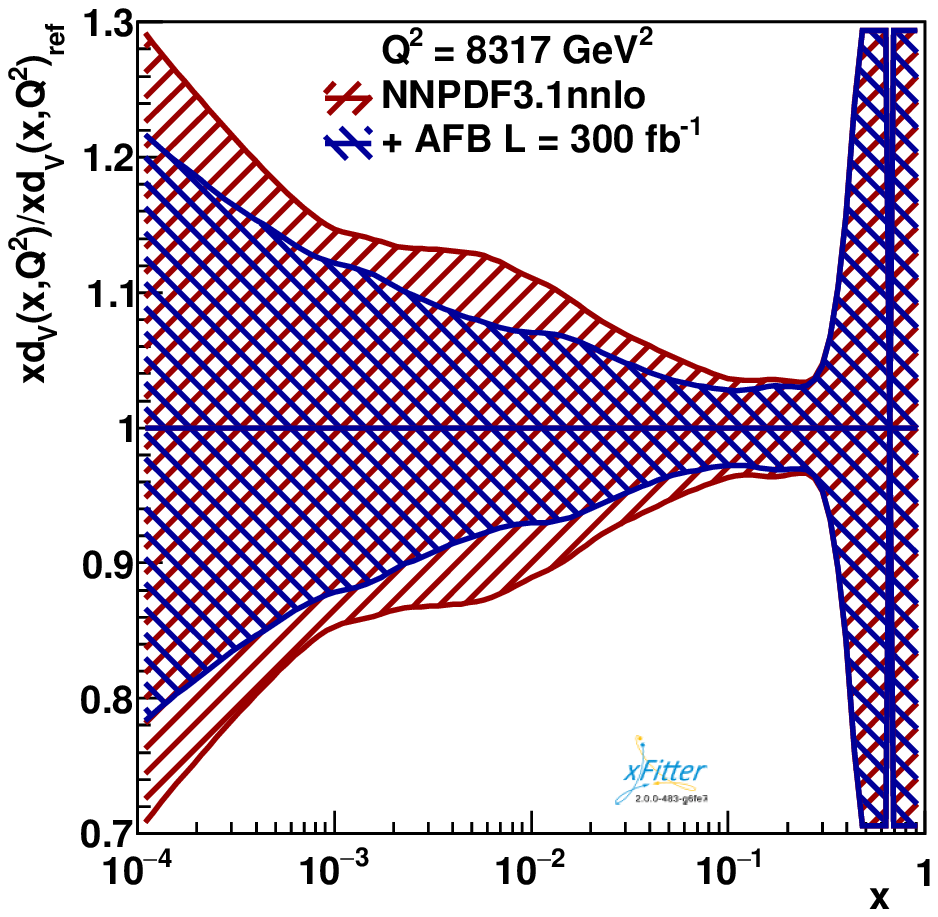}%
\includegraphics[width=0.25\textwidth]{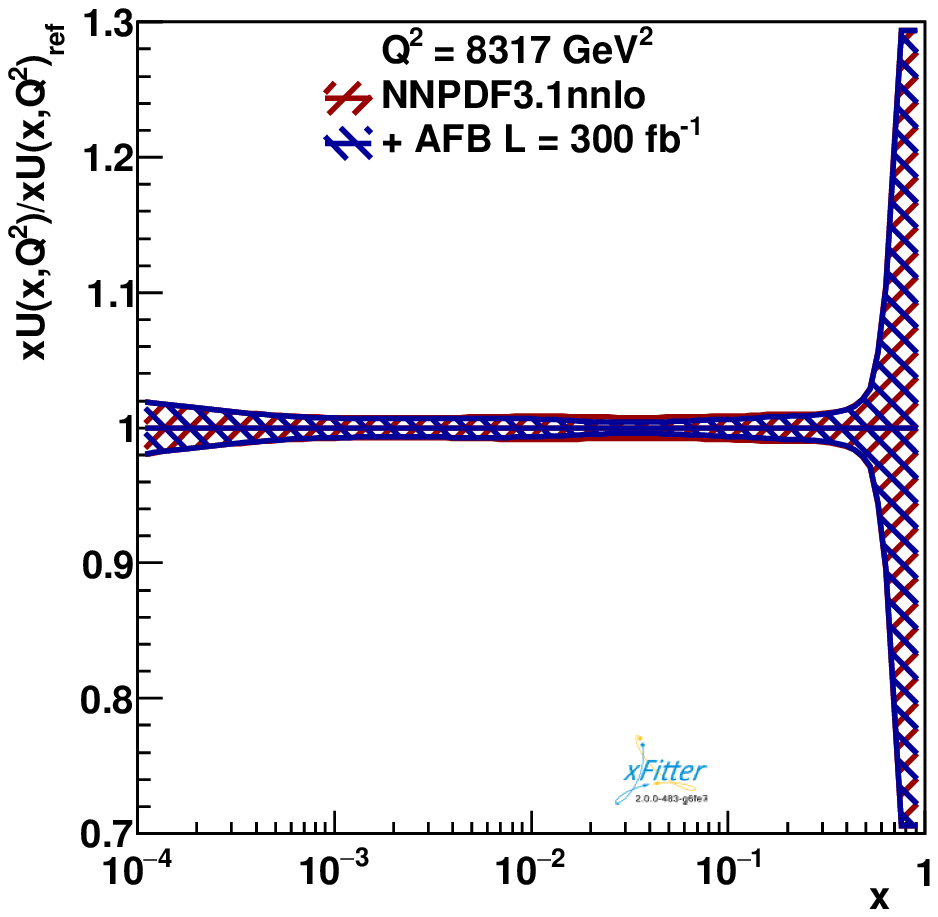}%
\includegraphics[width=0.25\textwidth]{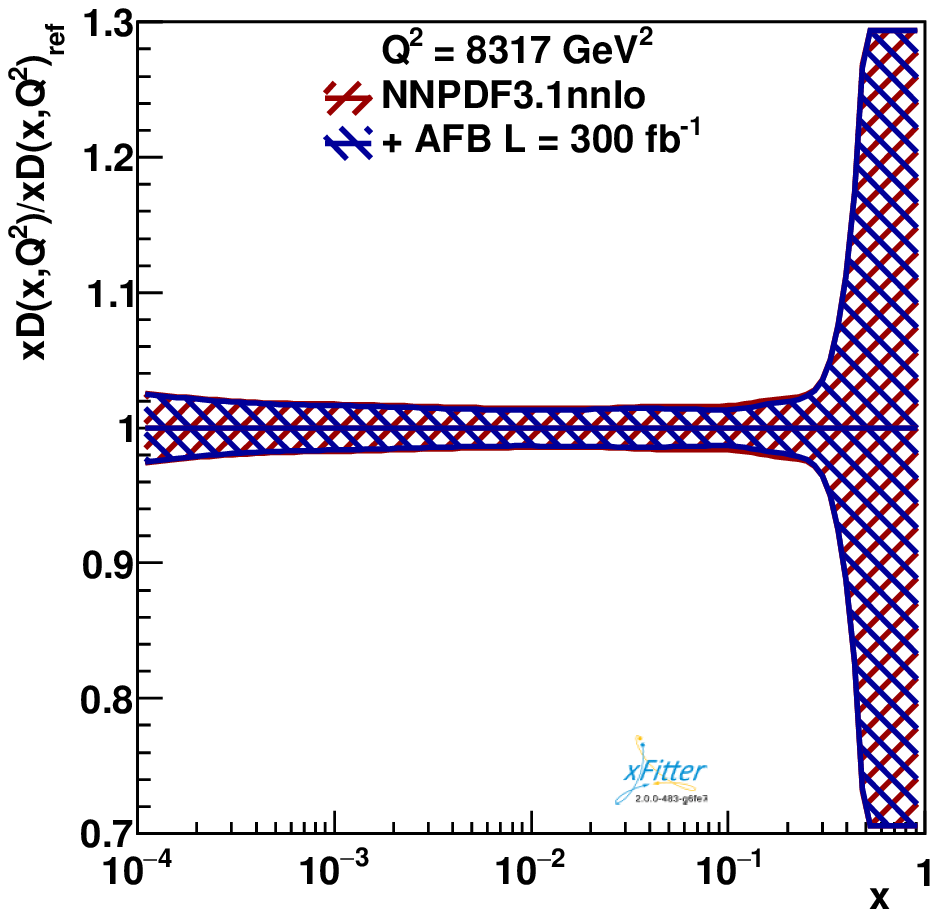}\\
\includegraphics[width=0.25\textwidth]{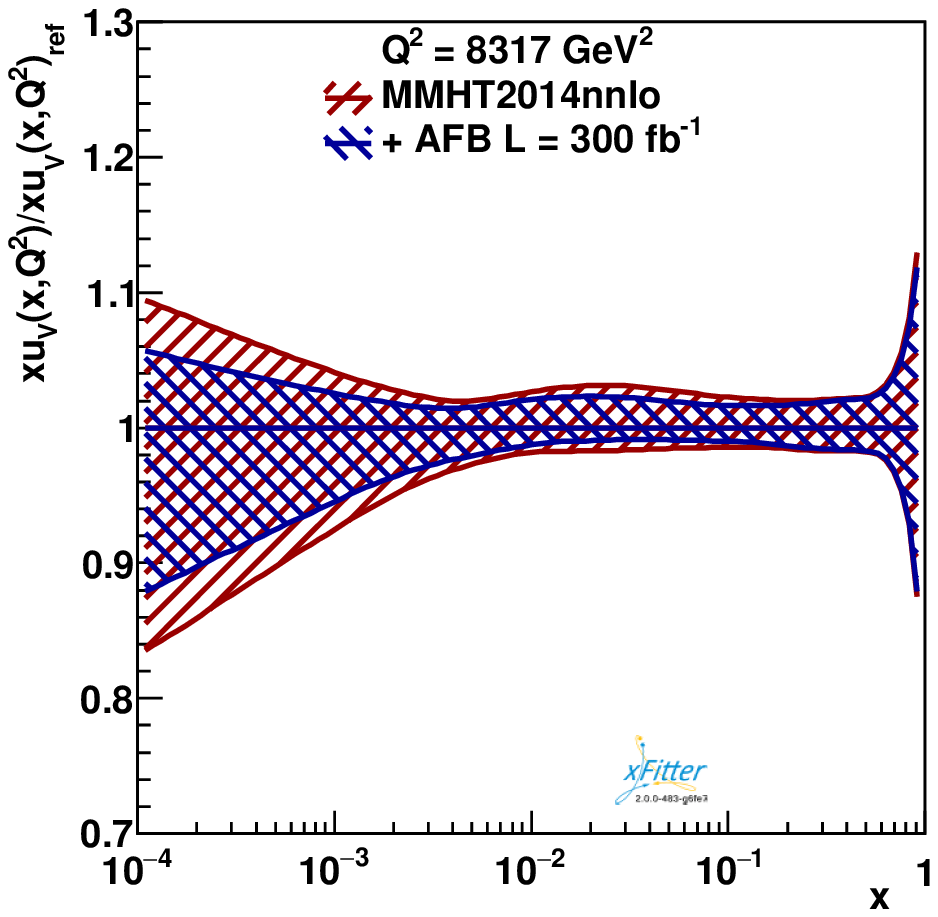}%
\includegraphics[width=0.25\textwidth]{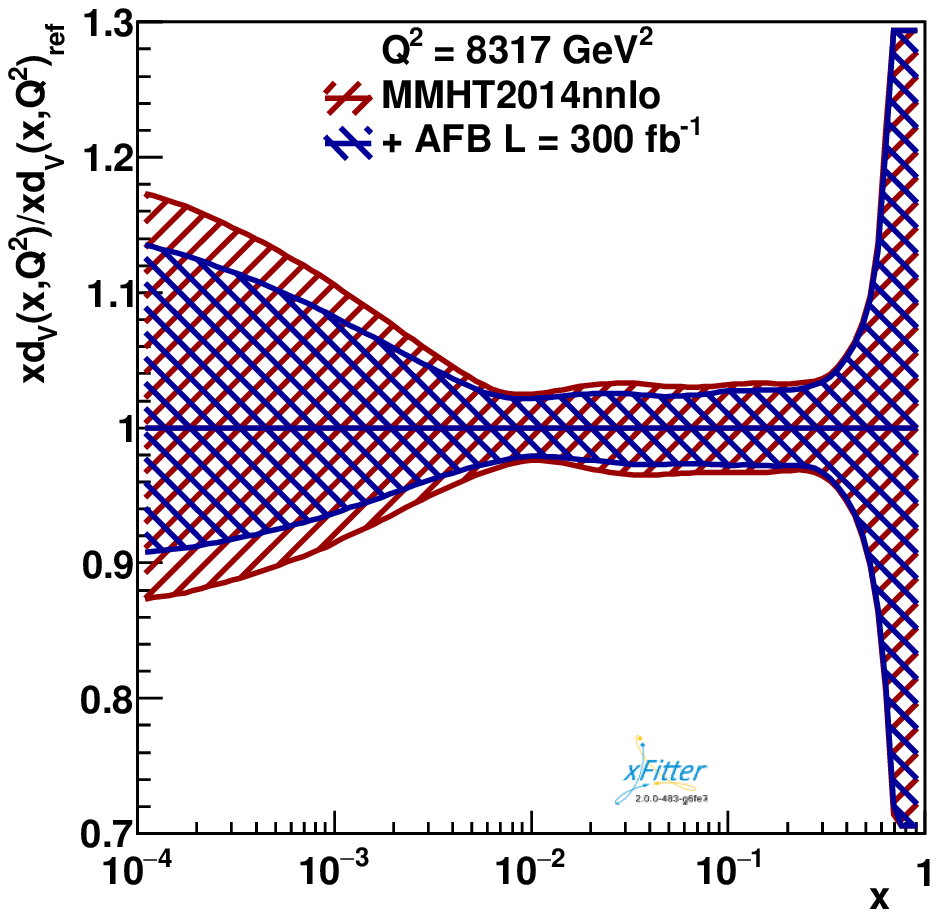}%
\includegraphics[width=0.25\textwidth]{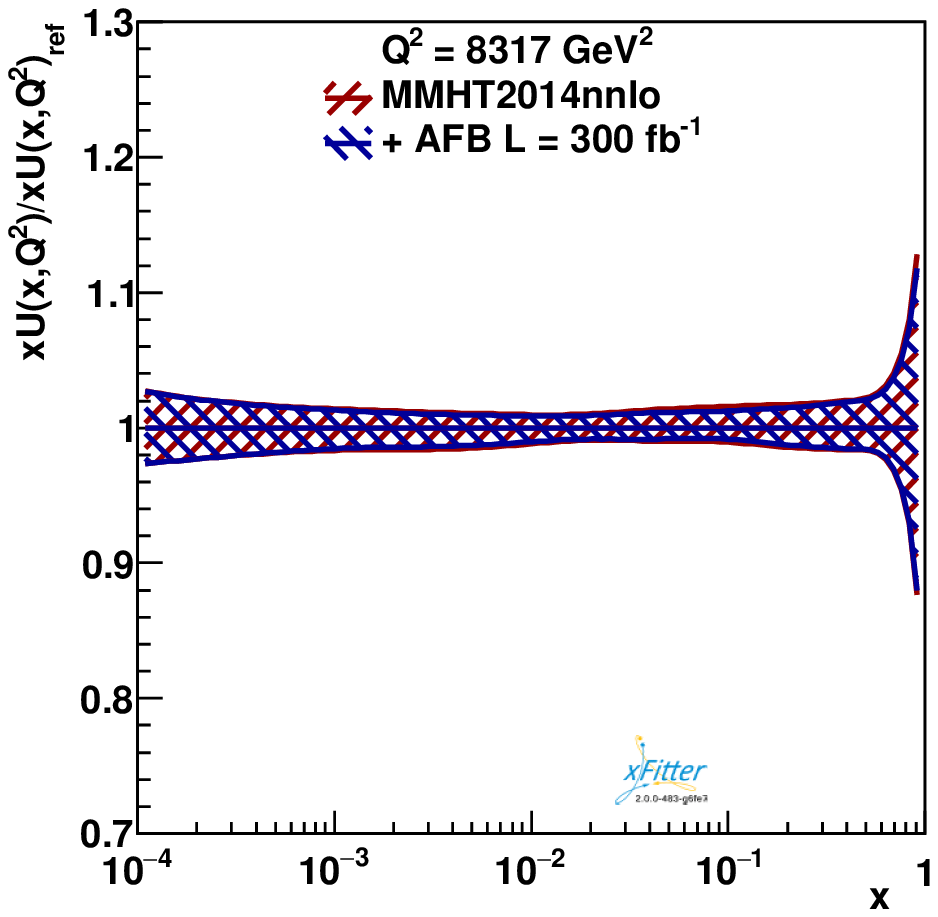}%
\includegraphics[width=0.25\textwidth]{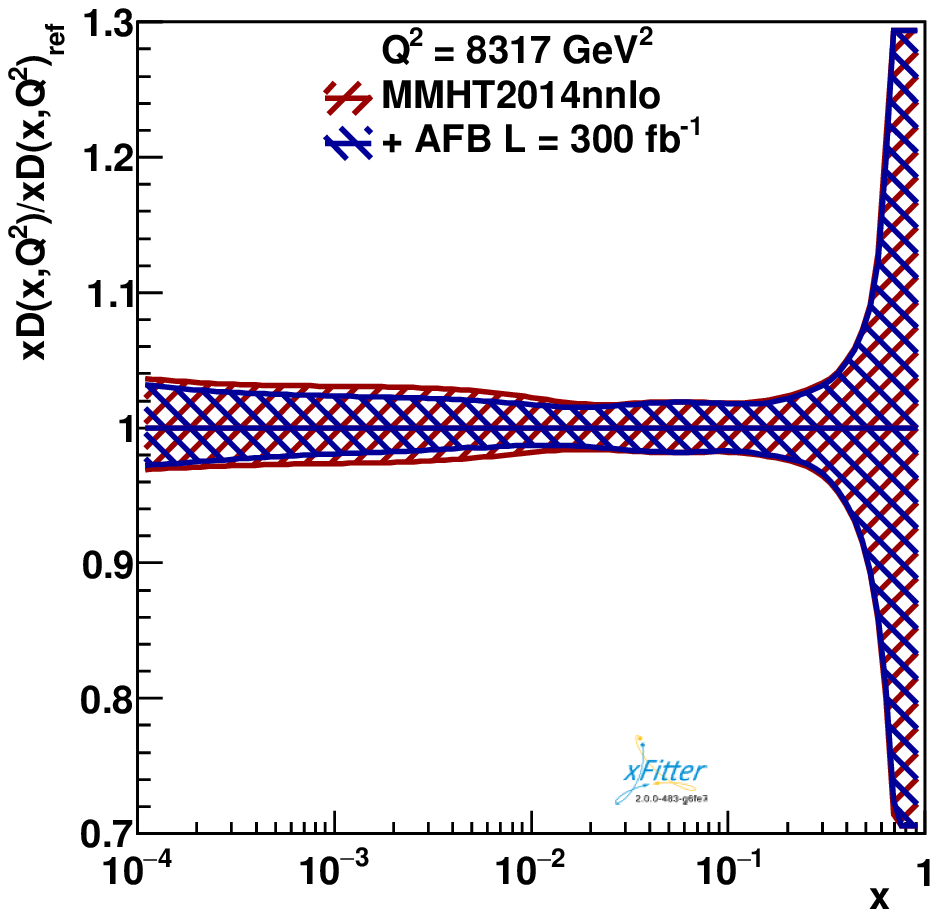}\\
\includegraphics[width=0.25\textwidth]{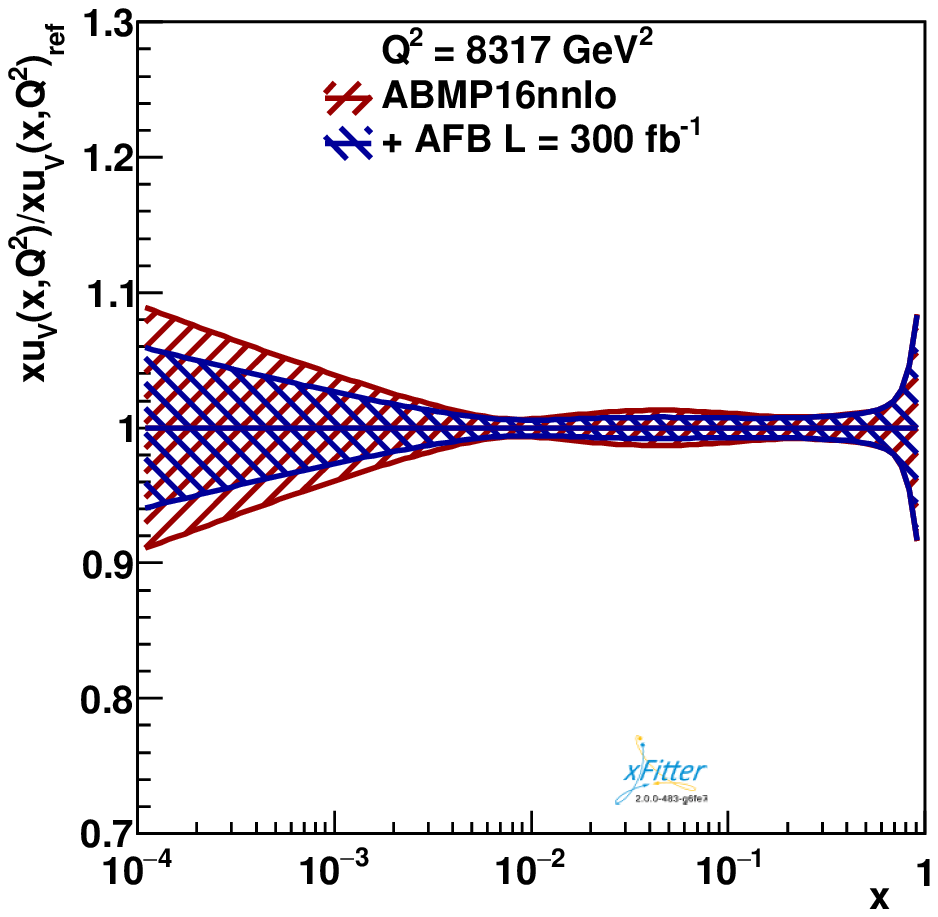}%
\includegraphics[width=0.25\textwidth]{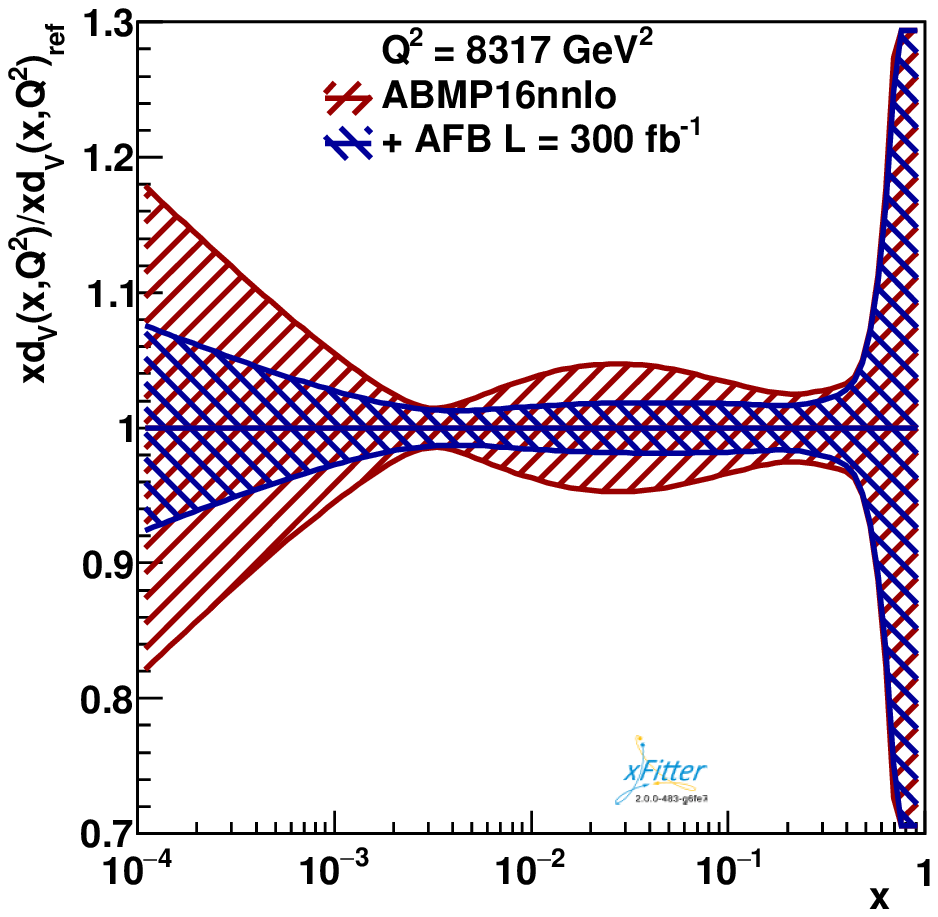}%
\includegraphics[width=0.25\textwidth]{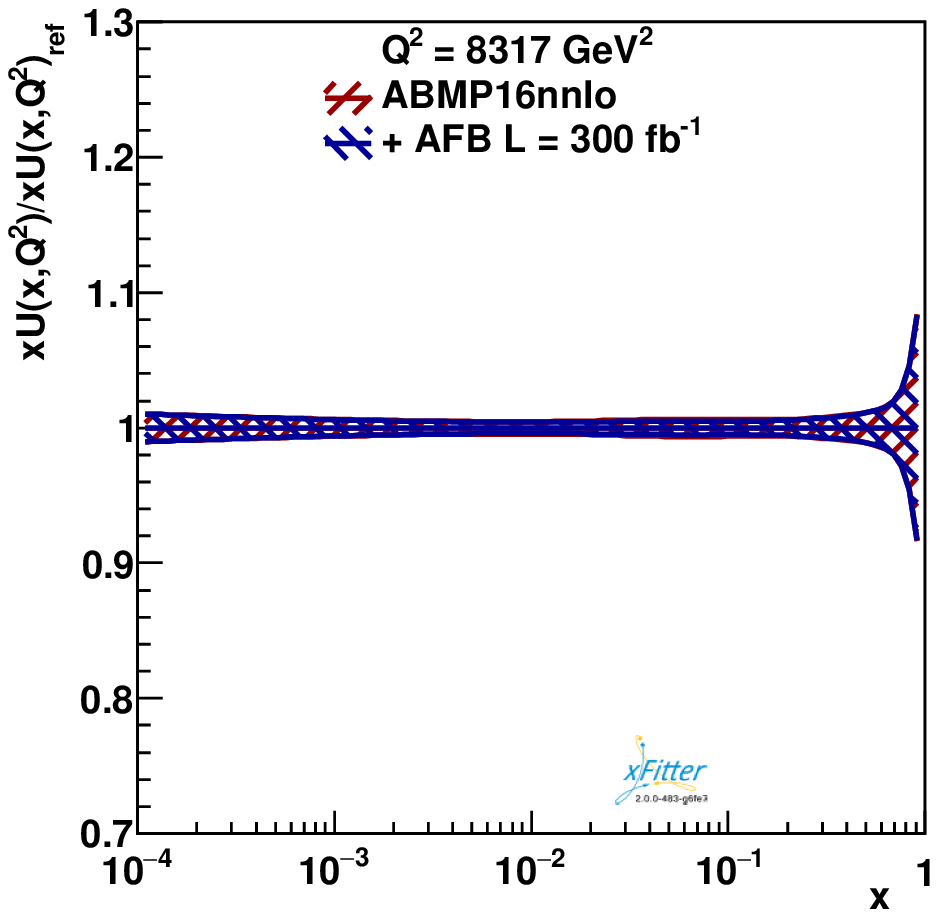}%
\includegraphics[width=0.25\textwidth]{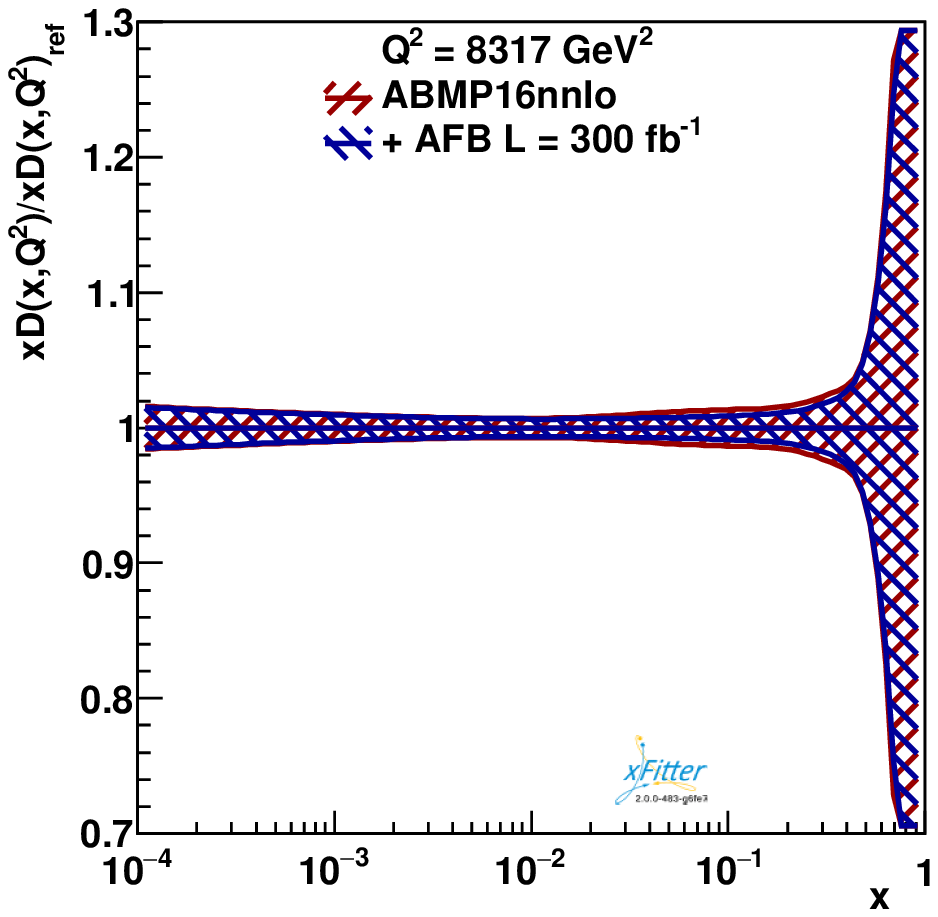}\\
\includegraphics[width=0.25\textwidth]{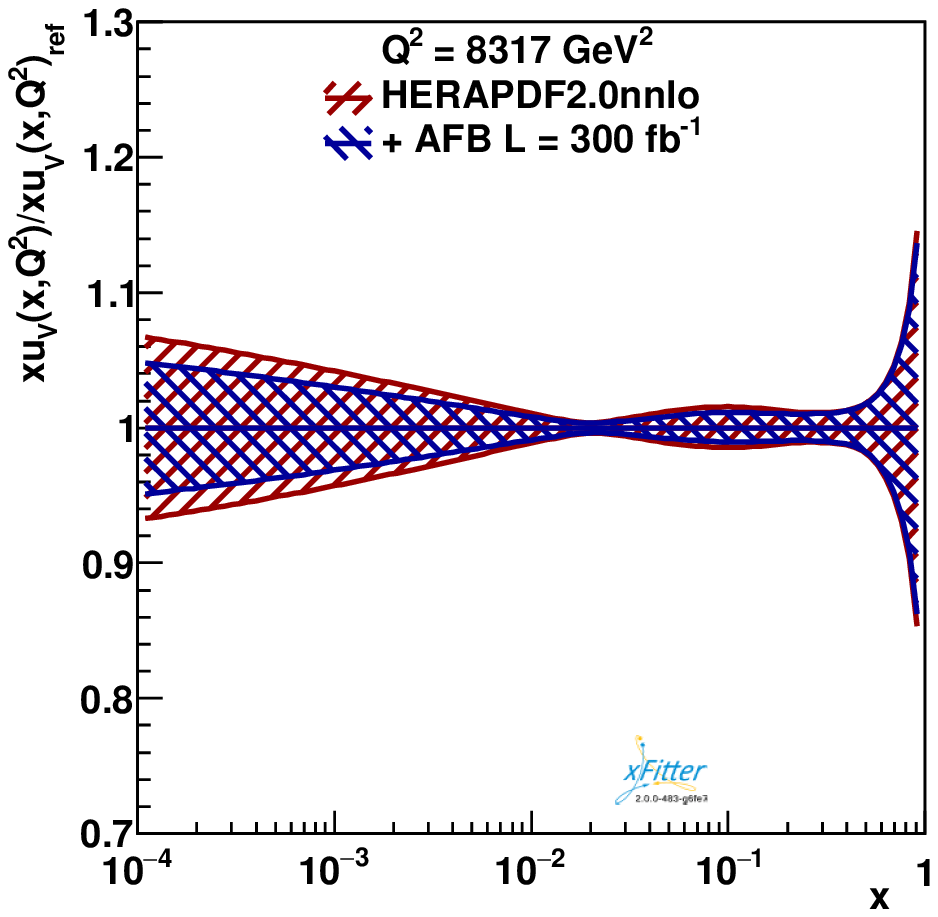}%
\includegraphics[width=0.25\textwidth]{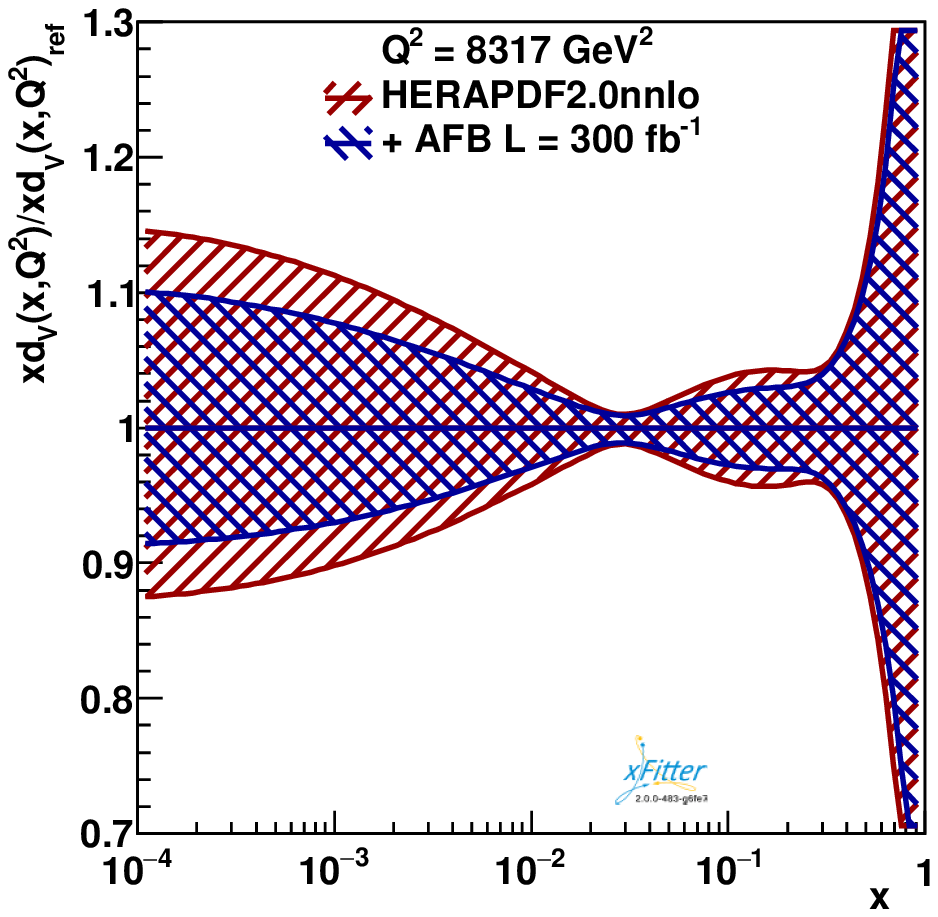}%
\includegraphics[width=0.25\textwidth]{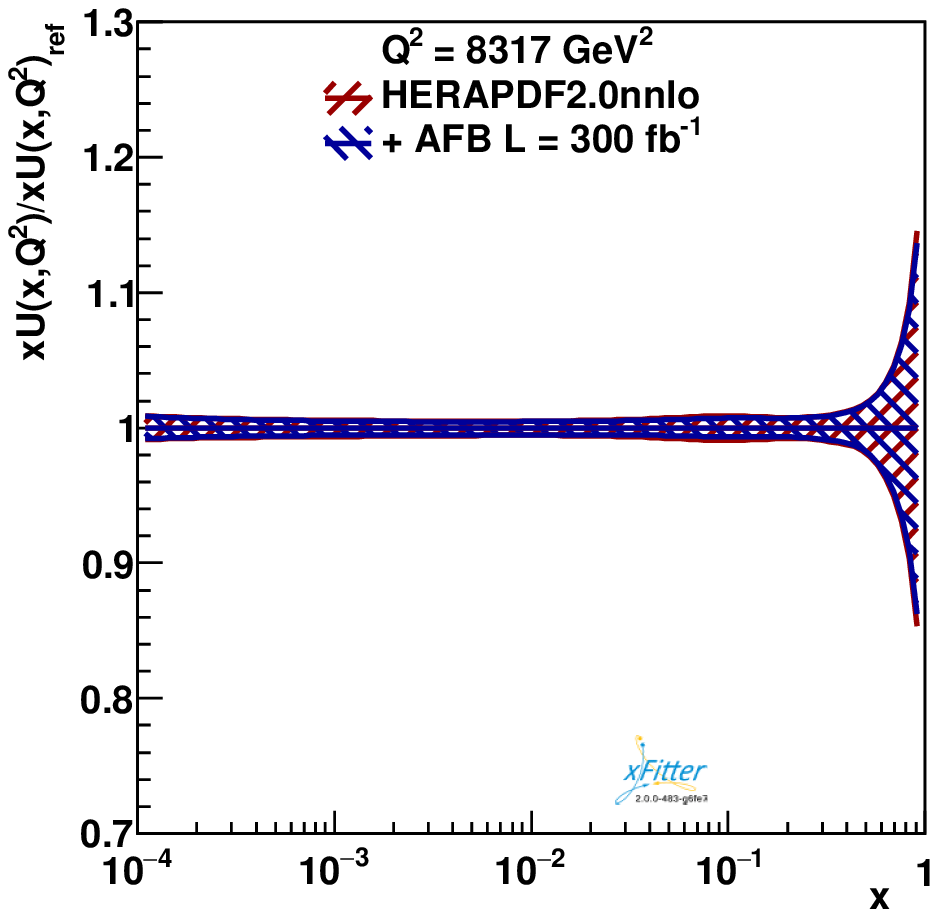}%
\includegraphics[width=0.25\textwidth]{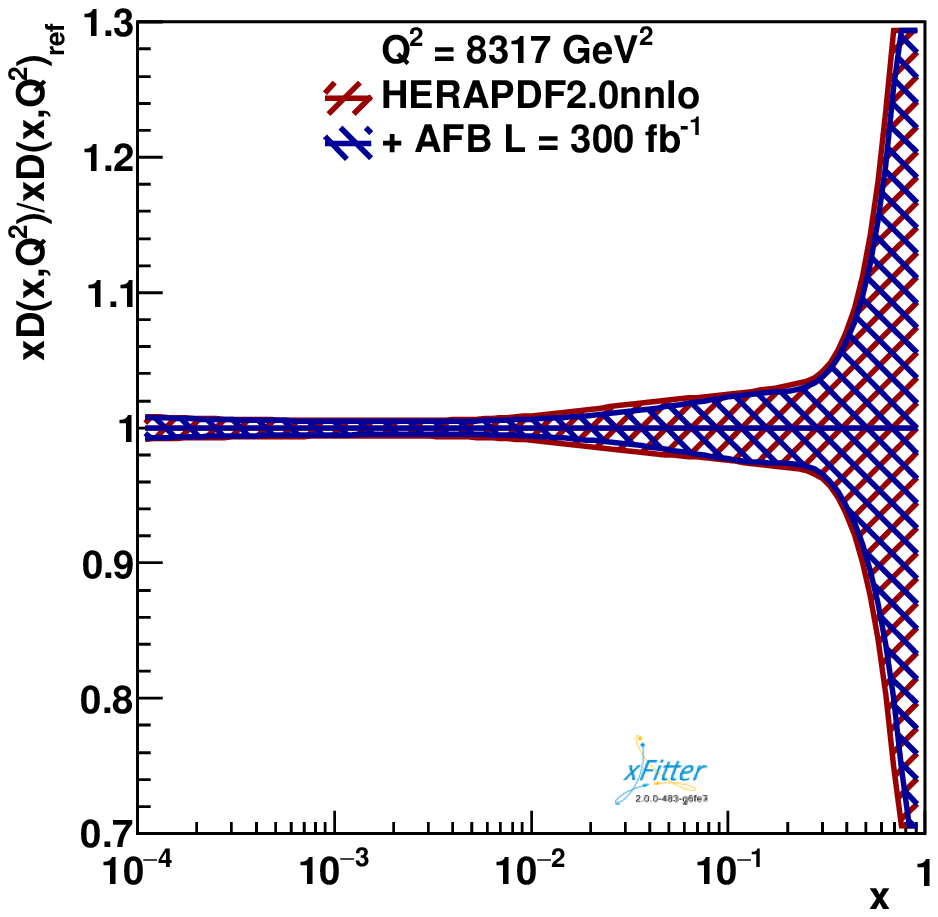}\\
\caption{Original (red) and profiled (blue) distributions for the normalised distribution of the ratios of (left to right) $u$-valence, $d$-valence, $u$-sea and $d$-sea quarks.
The profiled curves are obtained using $A_{\rm{FB}}$ pseudodata corresponding to an integrated luminosity of 300 fb$^{-1}$. Distributions are shown for the PDF sets (rows top to bottom) NNPDF3.1nnlo, MMHT2014nnlo, ABMP16nnlo and HERAPDF2.0nnlo.}
\label{fig:prof_PDFs}
\end{center}
\end{figure}

Figs.~\ref{fig:prof_CT14nnlo_lum} and \ref{fig:prof_PDFs} show results from the profiling analysis~\cite{Accomando:2019}, illustrating the 
 reduction  of  PDF uncertainties for various scenarios of $ A_{\rm{FB}} $ pseudodata and various PDF sets. 
In Fig.~\ref{fig:prof_CT14nnlo_lum} the role of different integrated luminosities is illustrated for the case of valence quark distributions in the 
CT15nnlo set~\cite{Dulat:2015mca}.  In Fig.~\ref{fig:prof_PDFs} the cases of valence and sea quark distributions are illustrated for 
NNPDF3.1nnlo~\cite{Ball:2017nwa}, MMHT2014nnlo~\cite{Harland-Lang:2014zoa}, 
ABMP16nnlo~\cite{Alekhin:2017kpj} and HERAPDF2.0nnlo~\cite{Abramowicz:2015mha},    
 using $A_{\rm{FB}}$ pseudodata corresponding to  integrated luminosity of 300 fb$^{-1}$. 
The largest effects are observed for $u$-valence  and $d$-valence distributions in the 
 region of intermediate and low momentum fraction $x$, and for ABMP16nnlo and HERAPDF2.0nnlo sets. 
Sea quark determinations show a   moderate improvement~\cite{Accomando:2019}, progressively increasing with the integrated 
 luminosity.  For  PDF sets with Hessian eigenvectors, it is shown explicitly  in~\cite{Accomando:2019} by eigenvector 
 reparameterization  that $u$-valence  and $d$-valence eigenvectors are highly correlated and $ A_{\rm{FB}} $ data 
constrain their charge-weighted sum $(2/3) u_V + (1/3) d_V$. 

 Ref.~\cite{Accomando:2019} also  studies different scenarios corresponding to different 
 selection cuts on the di-lepton rapidity.  
By increasing the rapidity cut, enhanced sensitivity is obtained to quark  distributions in  the high $x$ region.
In this case the high statistics of  the HL-LHC  is crucial  to achieve  
sufficient precision in the measurement of the $A_{\rm{FB}}$.

In summary,  the study reported in this article shows that  
neutral-current DY data from Run II, III and HL-LHC can be exploited 
to constrain nonperturbative QCD effects from PDFs, and thus to 
reduce the theoretical systematics affecting both precision 
SM studies and BSM searches. 
The $A_{\rm{FB}}$, in particular,  plays a complementary role to the lepton charge asymmetry of the 
DY charged-current channel, which has long been used in PDF global fits.  Traditionally the 
$A_{\rm{FB}}$ has been used for determinations of the weak mixing angle $\theta_W$. 
We have found that new PDF sensitivity arises from the di-lepton mass and rapidity spectra of the $A_{\rm{FB}}$, which 
encodes information on the lepton polar angle, or pseudorapidity. 
We have  presented quantitative results on   PDF uncertainties 
based on PDF profiling calculations in    {\tt{xFitter}}.  The results strongly support 
using DY data for combined determinations of $\theta_W$ and PDFs.

\section*{Acknowledgments}

Many thanks  to    the Moriond organizers and staff  for the invitation and for the pleasant atmosphere at 
 this  very  interesting   conference.

\end{document}